\documentstyle[11pt]{article}
\begin{document}
\title{Spectral geometry and quantum gravity}
\author{Giampiero Esposito\\
{\it INFN, Sezione di Napoli, 
Mostra d'Oltremare Padiglione 20, 80125 Napoli, Italy}}
\date{}
\maketitle
\begin{abstract}
Recent progress in quantum field theory and quantum gravity 
relies on mixed boundary conditions involving both normal and
tangential derivatives of the quantized field. In particular,
the occurrence of tangential derivatives in the boundary operator
makes it possible to build a large number of new local invariants.
The integration of linear combinations of such invariants of the
orthogonal group yields the boundary contribution to the 
asymptotic expansion of the integrated heat-kernel. This can be
used, in turn, to study the one-loop semiclassical approximation.
The coefficients of linear combination are now being computed
for the first time. They are universal functions, in that are
functions of position on the boundary not affected by conformal
rescalings of the background metric, invariant in form and 
independent of the dimension of the background Riemannian
manifold. In Euclidean quantum gravity, the problem arises of
studying infinitely many universal functions.
\end{abstract}

In classical and quantum field theory, as well as in the current
attempts to develop a quantum theory of the universe and of
gravitational interactions, it remains very useful to describe
physical phenomena in terms of differential equations for the
variables of the theory, supplemented by boundary conditions
for the solutions of such equations. For example, the problems
of electrostatics, the analysis of waveguides, the theory of
vibrating membranes, the Casimir effect, van der Waals forces,
and the problem of how the universe could evolve from an initial
state, all need a careful assignment of boundary conditions. 
In the latter case, if one follows a path-integral approach, one
faces two formidable tasks: (i) the specification of the geometries
occurring in the ``sum over histories" and matching the assigned
boundary data; (ii) the choice of boundary conditions on metric
perturbations which may lead to the evaluation of the one-loop
semiclassical approximation.

Indeed, while the full path integral for quantum gravity is a
fascinating idea but remains a formal tool, the one-loop 
calculation may be put on solid ground, and appears particularly
interesting because it yields the first quantum corrections 
to the underlying classical theory (although it is well known that
quantum gravity based on Einstein's theory is not perturbatively
renormalizable). Within this framework, it is of crucial importance
to evaluate the one-loop divergences of the theory under 
consideration. Moreover, the task of the theoretical physicist is
to understand the deeper general structure of such divergences.
For this purpose, one has to pay attention to all geometric 
invariants of the problem, in a way made clear by a branch of
mathematics known as invariance theory [1]. The key geometric 
elements of our problem are hence as follows.

A Riemannian geometry $(M,g)$ is given, where the manifold $M$
is compact and has a boundary
$\partial M$ with induced metric $\gamma$, and
the metrics $g$ and $\gamma$ are positive-definite. A vector
bundle over $M$, say $V$, is given, and one has also to consider
a vector bundle $\widetilde V$ over $\partial M$. An operator of
Laplace type, say $P$, is a second-order elliptic operator with
leading symbol given by the metric. Thus, one deals with a map
from the space of smooth sections of $V$ onto itself,
\begin{equation}\label{01}
P: C^{\infty}(V,M) \rightarrow C^{\infty}(V,M),
\end{equation}
which can be expressed in the form
\begin{equation}\label{02}
P=-g^{ab} \; \nabla_{a}^{V} \; \nabla_{b}^{V}-E,
\end{equation}
where $\nabla^{V}$ is the connection on $V$, and $E$ is an
endomorphism of $V$: $E \in {\rm {End}}(V)$. Moreover, the
boundary operator is a map
\begin{equation}\label{03}
{\cal B}: C^{\infty}(V,M) \rightarrow 
C^{\infty}({\widetilde V},{\partial M}),
\end{equation}
and contains all the informations on the boundary conditions
of the problem. Since we are interested in a generalization of
Robin boundary conditions [2--9], 
we consider a boundary operator
of the form (the operation of restriction to the boundary
being implicitly understood)
\begin{equation}\label{04}
{\cal B}=\nabla_{N}+{1\over 2}\Bigr[\Gamma^{i}
{\widehat \nabla}_{i}+{\widehat \nabla}_{i}\Gamma^{i}\Bigr]
+S .
\end{equation} 
With our notation, $\nabla_{N}$ is the normal derivative 
operator $\nabla_{N} \equiv N^{a}\nabla_{a}$ ($N^{a}$ being the
inward-pointing normal to $\partial M$), $S$ is an endomorphism
of the vector bundle $\widetilde V$, $\Gamma^{i}$ are
endomorphism-valued vector fields on $\partial M$,
and ${\widehat \nabla}_{i}$ 
denotes tangential covariant differentiation with respect to the
connection induced on $\partial M$. More precisely, when sections 
of bundles built from $V$ are involved, ${\widehat \nabla}_{i}$
means
$$
\nabla_{\partial M}^{({\rm lc})} \otimes 1
+ 1 \otimes \nabla ,
$$
where $\nabla_{\partial M}^{({\rm lc})}$ denotes the
Levi-Civita connection of the boundary of $M$. Hereafter, we
assume that $1+\Gamma^{2} >0$, to ensure strong ellipticity
of the boundary-value problem [9].

The case of mixed boundary conditions corresponds to the
possibility of splitting the bundle $V$, in a neighbourhood
of $\partial M$, as the direct sum of two bundles, say $V_{1}$
and $V_{2}$, for each of which a boundary operator of the 
Dirichlet or (generalized) Robin type is also given. The former
involves a projection operator, say $\Pi$, while the latter may
also involve the complementary projector, $1 - \Pi$, and the
metric of $V$, say $H$ [5]:
\begin{equation}\label{05}
{\cal B}_{1}=\Pi ,
\end{equation}
\begin{equation}\label{06}
{\cal B}_{2}=(1 - \Pi)\Bigr[H \nabla_{N}
+{1\over 2}\Bigr(\Gamma^{i}{\widehat \nabla}_{i}
+{\widehat \nabla}_{i}\Gamma^{i}\Bigr)+S \Bigr].
\end{equation}

We can now come back to our original problem, where only the
boundary operator (4) occurs, and investigate its effect on
heat-kernel asymptotics [8,9]. Indeed, given the heat
equation for the operator $P$, its kernel, i.e. the heat kernel,
is, by definition, a solution for $t > 0$ of the equation
\begin{equation}\label{07}
\left({\partial \over \partial t}+P \right)U(x,x';t)=0,
\end{equation}
jointly with the initial condition 
\begin{equation}\label{08}
\lim_{t \to 0}\int_{M}dx' \sqrt{{\rm {det}}g(x')} \;
U(x,x';t) \rho(x')=\rho(x) ,
\end{equation}
and the boundary condition
\begin{equation}\label{09}
\Bigr[{\cal B}U(x,x';t)\Bigr]_{\partial M}=0.
\end{equation}
The fibre trace of the heat-kernel diagonal, i.e.
${\rm {Tr}}U(x,x;t)$, admits an asymptotic expansion which 
describes the {\it local} asymptotics, and involves interior
invariants and boundary invariants. Interior invariants are
built universally and polynomially from the metric, the
Riemann curvature $R_{\; \; bcd}^{a}$ of $M$, the bundle
curvature, say $\Omega_{ab}$, the endomorphism $E$, and their
covariant derivatives. By virtue of Weyl's work on the invariants
of the orthogonal group, these polynomials can be found by using
only tensor products and contraction of tensor arguments [7,10]. 
The asymptotic expansion of the integral
\begin{equation}\label{010}
\int_{M}dx \sqrt{{\rm {det}}g} \; {\rm {Tr}}U(x,x;t) \equiv
{\rm {Tr}}_{L^{2}} \Bigr(e^{-tP}\Bigr),
\end{equation}
yields instead the {\it global} asymptotics. For our purposes,
it is more convenient to weight $e^{-tP}$ with a smooth function
$f \in C^{\infty}(M)$, and then consider the asymptotic 
expansion
\begin{equation}\label{011}
{\rm {Tr}}_{L^{2}}\Bigr(f e^{-tP} \Bigr)  \equiv 
\int_{M}dx \sqrt{{\rm {det}}g} \; f(x){\rm {Tr}} U(x,x;t) 
\sim (4\pi t)^{-m/2} \sum_{l=0}^{\infty}
t^{l/2}A_{l/2}(f,P,{\cal B}).
\end{equation}
Following Ref. [8], $m$ is the dimension
of $M$, and the coefficient $A_{l/2}(f,P,{\cal B})$ consists
of an interior part, say $C_{l/2}(f,P)$, and a boundary part,
say $B_{l/2}(f,P,{\cal B})$. The interior part vanishes for 
all odd values of $l$, whereas the boundary part only vanishes
if $l=0$. The interior part is obtained by integrating over $M$ the
linear combination of local invariants of the appropriate 
dimension mentioned above, where the coefficients of the linear
combination are {\it universal constants}, independent of $m$.
Moreover, the boundary part is obtained upon integration over 
$\partial M$ of another linear combination of local invariants. 
In that case, however, the structure group is $O(m-1)$ [10],
and the coefficients of linear combination are {\it universal
functions} [8], independent of $m$, unaffected by conformal  
rescalings of $g$, and invariant in form (i.e. they are 
functions of position on the boundary, whose form is independent
of the boundary being curved or totally geodesic). It is thus
clear that the general form of the $A_{l/2}$ coefficient is a
well posed problem in invariance theory, where one has to take
all possible local invariants built from $f,R_{\; \; bcd}^{a},
\Omega_{ab},K_{ij},E,S,\Gamma^{i}$ and their covariant
derivatives (hereafter, $K_{ij}$ is the extrinsic-curvature
tensor of the boundary), eventually integrating their linear
combinations over $M$ and $\partial M$. For example, in the
boundary part $B_{l/2}(f,P,{\cal B})$, the local invariants
integrated over $\partial M$ are of dimension $l-1$ in
tensors of the same dimension of the second fundamental form
of the boundary, for all $l \geq 1$ [1,10]. 
The universal functions
associated to all such invariants can be found by using the
conformal-variation method described 
in Refs. [1,7,8,10], jointly
with the analysis of simple examples and particular cases.

In other words, recurrence relations of algebraic nature exist
among all universal functions, and one can therefore use the
solutions of simple problems to determine completely the 
remaining set of universal functions for a given value of the
integer $l$ in the asymptotic expansion (11). 
The detailed investigation of the coefficients $A_{1}, A_{3/2}$
and $A_{2}$ when the boundary operator is given by Eq. (4) and
all curvature terms are non-vanishing is performed in Ref. [8].
One then finds the result (which holds for all integer values
of $l \geq 2$)
\begin{equation}\label{012}
A_{l/2}(f,P,{\cal B})={\widetilde A}_{l/2}(f,P,{\cal B})
+\int_{\partial M}{\rm {Tr}}\Bigr[a_{l/2}(f,R,\Omega,K,E,
\Gamma,S)\Bigr],
\end{equation}
where ${\widetilde A}_{l/2}(f,P,{\cal B})$ is formally analogous
to the purely Robin case, but replacing the universal constants 
in the boundary terms with universal functions, whereas 
$a_{l/2}$ is a linear combination of all local invariants of the
given dimension which involve contractions with $\Gamma^{i}$. Our
task is now to derive an algorithm for the general form of
$a_{l/2}$, since it helps a lot to have a formula that clarifies
the general features of a scheme where the number of new invariants
is rapidly growing. Indeed, from Ref. [8], we know that, in $a_{1}$,
only one new invariant occurs: $fK_{ij}\Gamma^{i}\Gamma^{j}$, 
whereas in $a_{3/2}$ 11 new invariants occur, obtained by 
contraction of $\Gamma^{i}$ with terms like (tensor indices are
here omitted for simplicity)
$$
fK^{2}, fKS, f{\widehat \nabla}K,f{\widehat \nabla}S,fR,
f\Omega,f_{;N}K.
$$
In $a_{2}$, the number of new invariants is 68: 57 involve
contractions of $\Gamma^{i}$ with terms like
$$
fK^{3},fK^{2}S,fKS^{2},fRK,f\Omega K, fEK, fRS, f \Omega S,
fK{\widehat \nabla}K, fS {\widehat \nabla}K,
$$
$$
fK{\widehat \nabla}S,fS{\widehat \nabla}S,
f{\widehat \nabla}{\widehat \nabla}K, 
f{\widehat \nabla}{\widehat \nabla}S,
f \nabla R, f \nabla \Omega, f \nabla E,
$$
10 local invariants involve contractions of $\Gamma^{i}$ with
contributions like
$$
f_{;N}K^{2},f_{;N}KS,f_{;N}{\widehat \nabla}K,
f_{;N}{\widehat \nabla}S,f_{;N}R, f_{;N} \Omega,
$$
and the last invariant is $f_{;NN}K_{ij}\Gamma^{i}\Gamma^{j}$ [8].
It is thus clear that the knowledge of all local invariants 
in $a_{l/2}$ plays a role in the form of $a_{(l+1)/2}$, and
one can write the formulae
\begin{equation}\label{013}
a_{1}=f\sum_{i=1}^{i_{1}}{\cal U}_{i}^{(1,1)} \;
I_{i}^{(1)},
\end{equation}
\begin{equation}\label{014}
a_{3/2}=f\sum_{i=1}^{i_{2}}{\cal U}_{i}^{(3/2,3/2)} \;
I_{i}^{(3/2)}+f_{;N}\sum_{i=1}^{i_{1}}
{\cal U}_{i}^{(3/2,1)} \; I_{i}^{(1)},
\end{equation}
\begin{equation}\label{015}
a_{2}=f\sum_{i=1}^{i_{3}}{\cal U}_{i}^{(2,2)} \; I_{i}^{(2)}
+f_{;N}\sum_{i=1}^{i_{2}}
{\cal U}_{i}^{(2,3/2)} \; I_{i}^{(3/2)} 
+f_{;NN}\sum_{i=1}^{i_{1}}{\cal U}_{i}^{(2,1)} \; 
I_{i}^{(1)}.
\end{equation}
With our notation, $i_{1}=1,i_{2}=10,i_{3}=57$, and
${\cal U}_{i}^{(x,y)}$ are the universal functions, where
$i$ is an integer $\geq 1$, $x$ is always equal to the order
$l/2$ of $a_{l/2}$, and $y$ is equal to the label of the
invariant $I_{i}^{(y)}$, which does not contain $f$ or
derivatives of $f$ and is of dimension $2y-1$ in $K$ or in
tensors of the same dimension of $K$.

These remarks make it possible to write down a formula which
holds for all $l \geq 2$:
\begin{equation}\label{016}
a_{l/2}(f,R,\Omega,K,E,\Gamma,S)=\sum_{r=0}^{l-2}f^{(r)}
\sum_{i=1}^{i_{l-r-1}}{\cal U}_{i}^{(l/2,(l-r)/2)}[\Gamma^{2}]
I_{i}^{(l-r)/2}[R,\Omega,K,E,\Gamma,S],
\end{equation}
where $f^{(r)}$ is the normal derivative of $f$ of order $r$
(with $f^{(0)}=f$), and square brackets are used for the 
arguments of universal functions and local invariants,
respectively. The equations (12) and (16) represent the
desired parametrization of heat-kernel coefficients with
generalized boundary conditions, provided that the $\Gamma^{i}$
are covariantly constant [8].

One has now to evaluate the universal functions in the
general formulae for $A_{3/2},A_{2},A_{5/2}$
and so on. For the coefficients $A_{3/2}$
and $A_{2}$, results of a limited nature are available in
Ref. [8], which show that all universal functions are generated
from $\sqrt{1+\Gamma^{2}}$ and ${1\over \sqrt{-\Gamma^{2}}}
{\rm {Artanh}} \sqrt{-\Gamma^{2}}$. 
Upon completion of this hard piece of work, one
could perform the evaluation of all universal functions for
$A_{5/2}(f,P,{\cal B})$ as well, possibly with the help of
computers. For this purpose, one has to combine 
the conformal-variation method with the analysis
of simpler cases. As shown in Refs. [8,9], one then obtains a
quicker and more elegant derivation of the coefficient
$A_{1}(f,P,{\cal B})$. There are thus reasons to expect that,
in the near future, all heat-kernel coefficients with 
generalized boundary conditions may be obtained via a computer
algorithm in a relatively short time. This adds evidence in
favour of the understanding of general mathematical structures
being very helpful in providing the complete solution of
difficult problems in physics and mathematics. In particular,
from the point of view of quantum field theory in curved 
manifolds, this would mean an entirely geometric understanding
of the first set of quantum corrections to the underlying
classical theory, with the help of invariance theory [1],
functorial methods [10] and computer programs.

In Euclidean quantum gravity, however, if one uses the de Donder
gauge-averaging functional, and if one requires invariance of
the whole set of boundary conditions under infinitesimal
diffeomorphisms on metric perturbations, one finds boundary 
operators of the kind (5) and (6), where the matrix $\Gamma^{2}$
commutes with $S$ {\it but not} with $\Gamma^{i}$ [8]. This
implies in turn that there exist infinitely many different tensors
of the type [11]
$$
T_{(m)}^{ij}(\Gamma^{l}) \equiv {\rm {Tr}}
\Bigr[\alpha_{(m)}(\Gamma^{2})\Gamma^{i}
\beta_{(m)}(\Gamma^{2})\Gamma^{j}\Bigr],
$$
which can contribute already to the integrand for 
$A_{1}(f,P,{\cal B})$, upon contraction with $K_{ij}$.
Thus, for the boundary operator given by the direct sum of
Eqs. (5) and (6), even the $A_{1}$ coefficient is unknown. 

One thus faces a highly non-trivial problem. On the one hand,
analytic results exist for the $A_{2}$ coefficient with boundary
operator (5) and (6) in the particular case of a portion of flat
Euclidean background bounded by a three-sphere [6,12]. Moreover,
it has been shown in Ref. [5] that the boundary operator given
by the direct sum of (5) and (6) leads to a symmetric operator
on metric perturbations. However, in the non-commuting case 
relevant for Euclidean quantum gravity, even the building blocks
of geometric invariants involving $\Gamma^{i}$ are unknown. This
is why it remains unclear how to write a general and unambiguous
formula for heat-kernel coefficients. The solution of such a
problem is of crucial importance in quantum gravity for the
following reasons:
\vskip 0.3cm
\noindent
(i) to improve the understanding of BRST invariant boundary
conditions [13];
\vskip 0.3cm
\noindent
(ii) to obtain an entirely geometric description of the one-loop
divergences in quantum gravity and quantum supergravity [6];
\vskip 0.3cm
\noindent
(iii) as a first step towards the quantization in arbitrary
gauges on manifolds with boundary;
\vskip 0.3cm
\noindent
(iv) to clarify the differences between Yang-Mills fields and
the gravitational field;
\vskip 0.3cm
\noindent
(v) to complete the application of the effective-action
programme to perturbative quantum gravity.

\section*{Acknowledgments}

The author is much indebted to Ivan Avramidi and Alexander
Kamenshchik for scientific collaboration on the topics 
described in this paper.

\end{document}